\documentclass[draft]{article}
\usepackage[psamsfonts]{amssymb}
\newcommand{\be}{\begin{equation}}
\newcommand{\ee}{\end{equation}}
\newcommand{\dd}{{\rm d}}
\newcommand{\DD}{{\rm D}}
\newcommand{\vek}[1]{{\bf #1}}
\newcommand{\sect}[1]{\\[\baselineskip] {\bf #1} \\ \\ \noindent}
\begin{document}
\begin{flushleft}
Published in: Found. Phys. Lett. (1999) vol. 12, no. 5, p.p. 427--439.
\end{flushleft}
\parskip 0pt \parindent 0pt
{ \bf
   EQUIVALENCE PRINCIPLE AND RADIATION \\
   BY A UNIFORMLY ACCELERATED CHARGE
   } 
  \\[\baselineskip]
\parindent 6em

  {\bf
         A. Shariati,
         M. Khorrami
  } 
{ \it
  \\[\baselineskip]

  Institute for Advanced Studies in Basic Sciences,

  P.O. Box 159, Zanjan 45195, Iran.

  Institute for Studies in Theoretical Physics and Mathematics

  P.O. Box 5531, Tehran 19395, Iran.
} 
\\[\baselineskip]

   Received 5 January 1999
\\[\baselineskip]
\noindent
     We address the old question of whether or not a uniformly
     accelerated charged particle radiates, and consequently,
     if weak equivalence principle is violated by electrodynamics.
     We show that {\em radiation} has different meanings; some
     absolute, some relative. Detecting photons or electromagnetic
     waves is {\em not} absolute, it depends both on the electromagnetic
     field and on the state of motion of the antenna. An antenna used
     by a Rindler observer does not detect any radiation from a uniformly
     accelerated co-moving charged particle.  Therefore, a Rindler observer
     cannot decide whether or not he is in an accelerated lab or in a
     gravitational field. We also discuss the general case.
\\[\baselineskip]
\parindent 3em
\sect{1. INTRODUCTION}
  An accelerated charged particle radiates.  Einstein's equivalence
  principle, on the other hand, tells us that the laws of physics
  in a constant gravitational field are the same as the laws of
  physics in an accelerated rocket.  It is tempting therefore to
  conclude that a charged particle located on the table of our
  lab radiates.  (Since according to the inertial freely falling
  observer, this charged particle is accelerating.)  However,
  using the Maxwell's equations coupled to a static gravitational
  field, one can show that it does not radiate.  It seems that either
  Einstein's equivalence principle is violated by electrodynamics,
  or a uniformly accelerated charged particle does not radiate.

  The problem of radiation of a uniformly accelerated charged particle
  has an interesting and controversial history.  In brief, based on a
  work by M. Born in 1909 [1], first W. Pauli [2] and
  then M. von Laue [3] gave arguments saying that such a charge
  does not radiate. Independently, G. A. Schott [4] derived
  the fields and concluded that such a charge radiates [5].
  (His work was discussed in more detail by S. M. Milner [6].)
  Then in 1949, D. L. Drukey [7] published a short note in favour
  of radiation. In 1955, M. Bondi and T. Gold [8] have asserted
  that the Born solution did not treat the singularity of the potentials
  on the light cone correctly.

  In 1960, T. Fulton and F. R\"ohrlich [9] published a paper
  discussing the problem in detail and concluding that: 1) ``If the
  Maxwell-Lorentz equations are taken to be valid, and we consider
  retarded potentials only, and if radiation is defined in the usual
  Lorentz invariant manner, a uniformly accelerated charge radiates
  at a constant non vanishing rate.'' 2) ``If one accepts the
  equations of motion based on the Abraham four-vector or on
  Dirac's classical electrodynamics, the radiation reaction
  vanishes, but energy is still conserved.'' 3) ``a charged
  and a neutral particle in a homogeneous gravitational field
  behave exactly alike, except for the emission of radiation
  from the charged particle.'' However, they argued, 4) ``Radiation
  is defined by the behaviour of the fields in the limit of large
  distance from the source. Correspondingly, an observer who wants
  to detect radiation {\em cannot do so in the neighbourhood} of
  the particle's geodesic.''

  In 1963 F. R\"ohrlich [10] concluded that a freely falling
  observer in a static gravitational field with vanishing Riemann
  tensor, would see a supported charge radiating, and vice versa,
  i.e. a supported observer would see a freely falling charge also
  radiating. R\"ohrlich's conclusions, which is in agreement with
  Einstein's principle of equivalence, was re-derived by a different
  and more transparent method by A. Kovetz and G. E. Tauber in 1969
  [11], and entered into the Problems of a text book by W. Rindler
  (problem 1.10 of [12]). In 1979, D. G. Boulware [13] wrote
  an article to show that ``the equivalence principle paradox that the
  co-accelerating observer measures no radiation while a freely falling
  observer measures the standard radiation of an accelerated charge is
  resolved by noting that all the radiation goes into a region of space
  time inaccessible to the co-accelerating observer''.

  In 1995, A. Singal [14] published a paper claiming again, and
  by a different method, that a uniformly accelerated charged particle
  does not radiate. This method and its conclusion is recently challenged
  by S.  Parrott [15].  Parrott somewhere else [16]
  argued that ``purely local experiments can distinguish a stationary
  charged particle is a static gravitational field from an accelerated
  particle in (gravity-free) Minkowski space'', in contradiction with
  Einstein's principle of equivalence.

  We see that, in addressing this problem some physicists argue that a
  uniformly accelerated charged particle does not radiate at all, while
  some other say that electrodynamics violates Einstein's equivalence
  principle; and it seems that the problem is still not resolved. In
  this article, we show that this paradox is due to a misuse of the
  word {\em radiation}.

  The scheme of the paper is as follows. In section II, we review
  the meaning of radiation.  In section III, we address the problem
  of detecting radiation of a uniformly accelerated charge by a
  Rindler observer, and show that no radiation is detected, simply
  because no magnetic field is measured.  In section IV, we show
  that this is due to a symmetry of the Minkowski spacetime:
  the symmetry with respect to Lorentz boosts, and we show that
  the same thing happens whenever the spacetime is static.
  We also show that in a stationary spacetime, there is an
  electromagnetic energy-like quantity, which is conserved for
  stationary charge distributions.  In sections V and VI, we
  address the question of radiation in terms of the world line
  of the particle. There, we conclude that a freely falling
  charge, or a charge whose four-velocity is proportional to
  a static time-like vector in a static spacetime, does not
  radiate.  Finally, section VII contains the concluding remarks.
\sect{ 2. RADIATION }
The word {\em radiation} reminds us three concepts:
\begin{enumerate}
  \item Flowing to infinity of energy in the form of electromagnetic waves.
  \item Detection of photons by a suitable device such as a photographic
        plate or an antenna.
  \item Deviation of the world-line of a charged particle from that
        of a neutral one of the same mass experiencing the same force.
\end{enumerate}
   The first notion is the one which is always used to define radiation in
   standard text books. The second notion seems to lead to the following
   argument: If a system radiates, one can detect it by counting photons. The
   energy or momentum of these photons may differ for different observers but,
   since the number of them must be the same, we conclude that radiation is in
   some sense absolute, i.e., whether or not a system radiates does not depend
   on the observer. (See for example p. 506 of [9].) This argument, which
   is the basis for the paradox mentioned, is false. The important point is
   that radiation has different notions, some are absolute some are relative.
   (The absolute notion is, in fact, the third one.)

   Consider a system of charged particles and the electromagnetic field
   produced by them. The divergence of the total energy-momentum tensor
   $T^{\mu \nu }=T_{{\rm e}}^{\mu \nu }+T_{{\rm m}}^{\mu \nu }$ vanishes.
   If $\xi ^\mu $ is a Killing vector field, $T^{\mu \nu }\xi _\mu $ is
   a conserved current. (Note that this is a local conservation law.)

   Conservation of the corresponding quantity depends only on the
   system (through $T^{\mu \nu }$) and the symmetry of the spacetime
   (through $\xi ^\mu $) and is not related to the observer or coordinate
   patch. In Minkowski spacetime the vector field $\partial _t$ is a
   Killing vector field which is {\em everywhere} time-like. Gauss'
   theorem then, leads to the standard definition of radiation as
   flowing energy to infinity by electromagnetic waves. This is an
   {\em absolute} notion of radiation, which is closely related to
   the third concept of radiation as well. But we must note that for
   a curved spacetime this argument may fail. For example, there may
   be no time-like Killing vector field; or there may be an event
   horizon which prevents us to enclose the charges with an sphere
   at infinity; etc. These difficulties make it sometimes impossible
   to define radiation as {\em flowing energy to infinity}.

   The second notion of radiation, viz. the detection of photons,
   is related to both the system and to the observer. Let $u^\mu $
   be the four-velocity of a local observer (not the source).
   This observer uses an apparatus to detect the electromagnetic
   flux of energy (or photons). Let $\sigma ^\nu $ be the space-like
   vector describing the direction of the apparatus used by him, and
   $T_{{\rm e}}^{\mu \nu }$ the energy-momentum tensor of the
   electromagnetic field. The amount of electromagnetic energy
   detected by the observer in proper time $d\tau $ is proportional
   to the area $d\Sigma $ of the apparatus used, and is equal to
   \begin{equation}
   dE=-T_{{\rm e}}^{\mu \nu }u_\mu \sigma _\nu d\Sigma d\tau .
   \end{equation}
   The quantity $\varepsilon:=-T_{{\rm e}}^{\mu \nu }u_\mu \sigma _\nu$
   is the flux of the electromagnetic energy through the surface
   $d\Sigma$ of the apparatus, and depends both on the system and
   on the world-line of the observer.

   Here it must be stressed that Fulton and R\"ohrlich's definition
   of radiation (referring to Synge [19]) is different [9, p. 506].
   To define radiation they use the quantity
   \begin{equation}
   I=T^{\mu \nu }v_\mu ^Qn_\nu
   \end{equation}
   where $v_\mu ^Q$ is the velocity four-vector of the {\em source} of
   radiation. (A minus sign is not here because they use different
   conventions.) Then, they look at this quantity at the null infinity.
   In doing this, one must transport $v_\mu ^Q$ away from the source
   location. This can be done only in Minkowski (flat) spacetime.
   In curved spacetimes, this is not a well defined quantity.
\sect{3. DISCUSSION OF THE PARADOX}
   Consider a charged particle $S$ located inside a rocket,
   which is uniformly accelerated with respect to an inertial
   frame I. (Specifically, we mean a Rindler rocket; see pp.
   49--51 and 156-157 of [12].) The question is whether
   or not the Rindler observer sees this charged particle
   radiating, i.e., whether or not he can detect photons.
   If he can, then he can deduce that he is inside an
   accelerating lab and not in a constant gravitational
   field. But we show that he cannot, and therefore,
   Einstein's principle of equivalence is not violated.
   Concretely speaking, we show:
   \begin{enumerate}
   \item If the Rindler observer uses a local static antenna,
         he will not  detect photons.
   \item There is a conserved current of the form
         $-T_{{\rm e}}^{\mu\nu}\Xi_\mu$, where $\Xi$ is the generator
         of translation in Rindler time. It is quite natural to call
         the  corresponding conserved quantity {\em the energy}.
   \end{enumerate}
   The proof of these two statements is based on a symmetry
   of the Rindler spacetime. Rindler frame is given by the
   following 3 parameter family of time-like world-lines:
   \begin{equation}
   x^2 - t^2 = X^2, \quad y = Y, \quad z= Z.
   \end{equation}
   For the spatial coordinate, the Rindler observer uses
   $(X,Y,Z)$, while for the time he uses the Lorentz boost
   parameter (or rapidity) $\omega := \tanh^{-1}t/x $.
   The proper time measured by a clock at $(X,Y,Z)$ is
   $\tau = \omega/X$. The world-line of an observer
   located at $(X,Y,Z)$ can be obtained by hyperbolic
   rotations in the $(t,x)$ plane. Rindler time $\omega$
   is simply the hyperbolic angle of this rotation.

   Now suppose there is a charged particle located at
   $(X_s,Y_s, Z_s)$ and an antenna at $(X_o, Y_o, Z_o)$.
   These objects are moving in the Minkowski spacetime.
   Since the Green's function for the 3+1 dimensional
   wave equation in Minkowski spacetime is non-zero
   only on the light-cone, we know that what $O$ measures
   at a Rindler time $\omega_o$ is due to the state of
   motion of $S$ at the retarded time, i.e., at the
   intersection of the past light-cone of the event
   $(\omega_o, X_o, Y_o, Z_o)$ with the world-line of $S$.

   To write the Poynting vector at Rindler instant
   $\omega_o$ for the local observer who is seated
   at $(X_o,Y_o, Z_o)$, we can write everything in
   the instantaneous rest frame of the source $S$
   at the retarded time and then use the Lorentz
   boost that transforms this frame to the
   instantaneous rest frame of $O$ (at the
   moment of observation).

   Since $S$ is uniformly accelerated, its acceleration
   at its instantaneous rest frame is $\alpha_s = 1 /X_s$.
   From electrodynamics we know that at the event
   $(t_o, x_o, y_o, z_o)$, the electromagnetic
   fields are given by
\begin{equation}
\vek{E} = q\left[ \frac{\vek{\hat{r}} - \vek{v}}{\gamma^2\left( 1 - \vek{v}
\cdot\vek{\hat{r}} \right)r^2}\right]_{{\rm ret}} + q \left[ \frac{%
\vek{\hat{r}} \times \left\{ \left( \vek{\hat{r}} - \vek{v}\right)\times
\dot{\vek{v}}\right\}}{\left( 1 - \vek{v}\cdot\vek{\hat{r}}\right)^3 r }
\right]_{{\rm ret}},
\end{equation}
\begin{equation}
\vek{B} = \left[ \vek{\hat{r}} \times \vek{E} \right]_{{\rm ret}}.
\end{equation}
   Here a hat means unit vector and the subscript ret means
   that everything must be computed at the retarded event
   $(t_s, x_s, y_s, z_s)$. Therefore, the electromagnetic
   field at the observation event is
\begin{equation}
\vek{E} = \frac{q}{r^2} + \frac{q\alpha}{r} \vek{\hat{r}}\times\left(
\vek{\hat{r}}\times\vek{\hat{x}}\right) ,
\end{equation}
\begin{equation}
\vek{B} = \frac{q\alpha}{r} \vek{\hat{x}}\times\vek{\hat{r}}.
\end{equation}
   It is important, however, to notice that these expressions
   are in the instantaneous rest frame of $S$.  To obtain
   electromagnetic fields as measured by the observer $O$,
   we must use the Lorentz transformations
\begin{equation}
\vek{E}^{\prime}= \gamma \left( \vek{E} + \vek{v}\times\vek{B}\right) -\frac{
\gamma^2}{\gamma +1}\vek{v}\vek{v}\cdot\vek{E},
\end{equation}
\begin{equation}
\vek{B}^{\prime}= \gamma \left( \vek{B} - \vek{v}\times\vek{E}\right) -\frac{
\gamma^2}{\gamma +1}\vek{v}\vek{v}\cdot\vek{B}.
\end{equation}
   We have chosen the inertial coordinate system such
   that at the retarded event the source's velocity
   $v_s = t_s / x_s$ vanishes; therefore, at the retarded
   event $t_s =0$. The world-line of the source is
   $x_s^2 - t_s^2 = X_s^2 = \alpha_s^{-2} = {\rm constant}$.
   From this it follows that the acceleration of the source
   at the retarded time is $\alpha = 1 / x_s$.

   From $x_o^2 - t_o^2 = X_o^2 = \alpha_o^{-2} = {\rm constant}$,
   which describes the world-line of the antenna
   (i.e. local observer), it follows that $v= t_o / x_o$.
   We also note that $r = (t_o - t_s) = t_o$ and
   $ \vek{\hat{x}}\cdot \vek{r} = x_o - x_s$.
   From these ingredients, it is easy to see that
\begin{equation}
\vek{B'} = 0.
\end{equation}
   This shows that a local Rindler observer $O$,
   whose world-line is given by
\begin{equation}
x_o^2 - t_o^2 = X_o^2, \quad y_o = Y_o, \quad z_o = Z_o,
\end{equation}
   sees a pure electric field and, therefore, no radiation.

   It is worthy of mention that Pauli's argument,
   based on the fields derived by Born, is also the
   vanishing of the magnetic field. However, as
   mentioned by Bondi and Gold [8], and Fulton
   and R\"ohrlich [9], the basis of his derivation
   is not true. Here again we see that the magnetic field
   vanishes for the (Rindler) co-moving observer, and we
   see this by exactly computing the fields as measured by
   this observer.

   The argument given above depends deeply on a symmetry
   of the Minkowski spacetime, viz. the existence of the
   time-like Killing vector field $t\partial_x + x\partial_t$,
   which for the Rindler observer is just $\partial_\omega$.
   In the next section we discuss the general case.
\sect{4. RADIATION IN TERMS OF THE ELECTROMAGNETIC \\
         ENERGY-MOMENTUM TENSOR}
   A stationary spacetime is a spacetime with a time-like
   Killing vector $\xi :=\partial /\partial t$.  In such a
   spacetime, one can chose a coordinate system in which,
   the metric components are $t$
   stationary charge distribution $J^\mu $ in a stationary
   spacetime: $\partial _0J^\mu =0$, and
   $\partial _\mu \left( \sqrt{|g|} J^\mu \right) =%
   \partial_i\left( \sqrt{|g|}J^i\right) =0$. Using
   a $t$-independent ansatz for the four-potential
   $A_\mu $, it is easily seen that the field-strength
   tensor $F_{\mu \nu }$ is $t$-independent. Moreover,
   the source-full Maxwell equation
\be \label{4}
    {\frac 1{{\sqrt{|g|}}}}\left(g^{\mu \alpha }g^{\nu \beta }
    F_{\alpha \beta }\sqrt{|g|}\right) _{,\nu}=J^\mu ,
\ee
   shows that there is no inconsistency in taking
   $F_{\mu \nu }$ independent of $t$, because both
   $j^\mu $ and $g_{\mu \nu }$ are $t$-independent.
   In fact, if $F_{\mu \nu }(t,{\bf r})$ is a solution
   to (\ref{4}), $F_{\mu \nu }(t+\Delta ,{\bf r})$ is
   also a solution. So, if the Maxwell equations have
   a unique solution in this spacetime, the field-strength
   tensor should be $t$-independent.

   Since $\xi $ is a Killing vector, we have
\be
     \left( -T^{\mu \nu }\xi _\nu \right) _{;\mu }=
     {\frac 1{{\sqrt{|g|}}}}\left( -\sqrt{|g|}T^\mu {}_0\right) _{,\mu }=0.
\ee
   So, a conserved current ${\cal J}^\mu :=-T^\mu {}_0$ exists.
   One can define a corresponding current for the electromagneti
   c part of the energy-momentum tensor. However, as
   $T^{\mu\nu}_{{\rm e}}$ is $t$-independent, it is seen that
\be
   \int_\Sigma \left( -T^\mu{}_0\right) _{{\rm em}}\sqrt{|g|}\dd S_\mu =
   \int_{\Sigma'}\left( -T^\mu{}_0\right) _{{\rm em}}\sqrt{|g|}\dd S_\mu ,
\ee
   where $\Sigma $ is any hyper-surface, and
   $\Sigma ^{\prime }$ is the hypersurface formed
   by translating $\Sigma $ along the Killing vector
   field $\xi $ by some value $\Delta $. This means
   that the energy-like quantity of the electromagnetic
   field in any hypersurface (any portion of space),
   including or excluding the charge(s), does not change.
   In this sense, one can say that a stationary charge
   distribution in a stationary spacetime does {\em not} radiate.

   A static spacetime is a stationary spacetime,
   for which there exists a family of space-like
   hyper-surfaces normal to the time-like Killing
   vector field $\xi $. This means that there exists
   a suitable choice of coordinates, for which
   $g_{0i}=g^{0i}=0$, and $g_{\mu \nu }$'s are
   all $t$-independent. In such a spacetime,
   consider a static charge distribution.
   A static charge distribution is a stationary
   one with the additional condition $J^i=0$.
   For the field produced by a static distribution
   one can take the ansatz: $A_i=0$ and $\partial _0A_0=0$.
   From this, and the fact that the metric is
   $t$-independent and block-diagonal, it is
   easy to see that
\be \label{11}
    F_{ij}=F_i{}^j=F^{ij}=0.
\ee
   This shows that the source-full Maxwell equation
   is identically satisfied for $\mu \ne 0$. For $\mu =0$,
\be
   -{\frac 1{{\sqrt{|g|}}}}
   \left( g^{00}g^{ij}\sqrt{|g|}\partial _jA_0\right) _{,i}=J^0.
\ee
   This equation is consistent, since the left-hand
   side is $t$-independent, as well as the right-hand side.

   This solution to the static charge distribution has no
   magnetic field, and has a $t$-independent electric field.
   By magnetic field (in a covariant form) we mean the tensor
   $B_{\mu\nu}:=F_{\alpha\beta}h^\alpha{}_\mu h^\beta{}_\nu$.
   Here $h_{\mu\nu}$ is the projector normal to $\xi$, that is
   $h_{\mu\nu}:=g_{\mu\nu}-%
   \left(\xi_\mu\xi_\nu\right)/\left(\xi\cdot\xi \right)$.
   Note that for the choice of coordinates introduced above,
   $h_{ij}=g_{ij}$, $h_{00}=h_{i0} = 0$, $B_{ij}=F_{ij}$,
   and $B_{0i}=0$. The electric field is similarly defined
   through $E_\mu :=F_{\mu\nu}\xi^\nu /\sqrt{-\xi\cdot\xi}$.

   From (\ref{11}), we have $-T_{{\rm e}}^i{}_0=%
   -h^\mu{}_\nu T_{{\rm e}}^\nu{}_\alpha\xi^\alpha =0$.
   Consider an observer with the four-velocity
   $u^\mu =\xi^\mu/\sqrt{ -\xi\cdot\xi}$.
   For this observer the magnetic field is
   just $B_{\mu\nu}$, and the electric field
   is $E_\mu$. So, this observer measures a
   $t$-independent electric field and no
   magnetic field. Moreover, what this observer
   measures as the Poynting vector is
\begin{eqnarray}
S^\mu &:=&-u_\nu T^{\alpha\nu}h^\mu{}_\alpha\nonumber\\
 &=&-u_\nu F^{\beta\nu}F_\beta{}^\alpha h^\mu{}_\alpha\nonumber\\
 &=&E^\beta B^\mu{}_\beta ,
\end{eqnarray}
   which is identically zero. This means that, in terms
   of the Poynting vector, this observer measures no
   radiation, simply because the magnetic field is zero
   for this observer. This conclusion is stronger than
   that of the last subsection. That meant no net flux
   of the electromagnetic {\em energy} is observed by
   the stationary observers. This means that, besides,
   no electromagnetic {\em energy} current is observed
   by such an observer.
\sect{5. RADIATION AND THE WORLD-LINE OF CHARGED \\ PARTICLES}
   Is it true that, in a gravitational filed, charged particles
   fall the same as uncharged particles? This problem is also
   related to the paradox mentioned at the beginning of this paper.

   Of course, this question must be answered experimentally
   (and the experiment is more difficult than it seems, cf.
   [18].) But let us study the answer given by the
   known theory of electrodynamics. The question is not
   trivial, for acceleration causes radiation and this
   may causes a damping. It seems therefore that a charged
   particle does not fall the same as an uncharged particle.
   To answer this question we have to know the form of the
   radiation reaction force. The best candidate for this,
   is the Lorentz-Abraham-Dirac force from which it follows
   that charged particles fall the same as neutral particles,
   i.e. the weak equivalence principle is fulfilled. To our
   knowledge, this was first noticed by R\"ohrlich [10].
   Here, we present a heuristic argument in favour of the
   conclusion that equivalence principle is not violated
   by charged particles.

   To begin with, let us consider a stationary charged
   particle in the Minkowski spacetime.  The world-line
   of such a particle is
\begin{equation}
-\infty <t_s<\infty ,\quad x_s={\rm const.}>0,\quad y_s=z_s=0.
\end{equation}
   A Rindler observer sees only the segment $-x_s<t<x_s$;
   by the following world-line:
\begin{equation}
-\infty <\omega _s<\infty ,\quad X_s=x_s/ \cosh \omega _s,\quad Y_s=Z_s=0.
\end{equation}
   Trivially, this world-line is a geodesic describing
   the motion of the particle approaching the horizon
   as $\omega \to \pm \infty $. Now let's interpret the
   Rindler spacetime as a gravitational field. Since we
   know that in the Minkowski spacetime the charged
   particle follows a geodesic, a little reflection
   shows that in the Rindler gravitational field a
   free charged particle falls the same as a free
   uncharged particle. To this problem, let's apply
   Einstein's equivalence principle. In the comoving
   freely falling lab, which is a local inertial frame,
   the charged particle is at rest and therefore it does
   not radiate. The comoving observer sees no reason for
   the charged particle to move in the lab, simply because
   it is completely free. Therefore, with respect to the
   freely falling lab, the charged particle is always at
   the same position. Transforming this result to the
   Rindler frame, we conclude that in the gravitational
   field of the Rindler spacetime, charged particles
   fall the same as the uncharged particles.
   The electromagnetic field produced by this
   charged particle as seen by the inertial
   comoving observer is purely electric, a
   Rindler observer, however, sees also a
   magnetic field and a non-vanishing Poynting
   vector; and sees that the charged particle
   goes to the horizon $X=0$ as $\omega\to\infty$.
   If the Rindler observer uses a local device, such
   as a camera, he will observe some photons, i.e. he
   will receive some energy which causes an effect on
   his photographic plate. This effect, however, is not
   radiation. To convince, suppose a charged particle
   is in uniform rectilinear motion relative to an
   inertial observer. If this observer uses an antenna,
   his antenna does receive some energy, simply because
   the Poynting vector at the position of antenna is
   non-zero (and it is even time varying).  However,
   this is simply the result of an interaction of
   antenna with the moving charge, (and of course,
   as a result of this interaction the charged
   particle's trajectory is affected).

   In the previous sections, it was shown that a uniformly
   accelerated charge in a Minkowski spacetime does {\em not}
   radiate, in the sense that for the Rindler observer the Poynting
   vector vanishes, and an energy-like quantity for the electromagnetic
   field is constant. This means that, according to Rindler observers,
   no extra force is needed to maintain the uniform acceleration of
   such a charged particle (of course no extra force beside the force
   needed for a neutral particle of the same mass to have that
   acceleration). In other words, the world-line of the charged
   particle will be the same as that of a neutral particle.

   Now consider a Rindler gravitational field, in which a charged
   particle is stationary. One can obtain the electromagnetic field
   of this particle in a manner exactly the same as that of section
   III, which shows that there is no radiation.

   Do, in some sense, these two problems differ?
   Some authors [16] argue that to support
   a charged particle in a gravitational field a rocket
   is needed, and this rocket spends more fuel than a
   rocket needed to do the same thing for a neutral particle.

   However, the results of previous sections, in terms
   of the electromagnetic energy, make no difference
   between charged and neutral particles in any of the
   two cases. According to Rindler observers, no extra
   force is needed for the charge, and as the four-vector
   of force is zero according to one observer, it should
   be zero according to other observers as well.  So the
   result of the above gedanken experiment should be null.

   Another problem is that of a freely falling charge
   in a Rindler gravitational field. Such a charge moves
   uniformly according to Minkowski observers, so that it
   does not radiate according to them, and its world-line
   should be the same as that of a neutral particle.
   In fact, this problem, in terms of energy considerations,
   is the same as the problem of a stationary (or uniformly
   moving) charge in Minkowski spacetime. These results are
   also in agreement with Einstein's equivalence principle.
\sect{6. RADIATION REACTION FORCE}
   We show that these results are also true in terms of the damping force,
   so that there is no rocket paradox. We show that this force is zero in
   certain cases, which are the generalisations of the above cases.

When a particle radiates, it experiences a force due to its radiation. (The
original derivation of the reaction force is due to Dirac [17]. A
recent derivation is given by A. Gupta and T. Padmanabhan [20].) The
relativistic form of the Abraham-Lorentz-Dirac force, experienced by an
accelerated charge is
\be \label{22}
    f_{{\rm ALD}}^\mu ={\frac{{e^2}}{{6\pi }}}\left(
    {\frac{{\DD^2u^\mu }}{{\DD\tau ^2}}}+u^\mu u_\alpha
    {\frac{{\DD^2u^\alpha }}{{\DD\tau ^2}}}\right) ,
\ee
where $e$ is the charge of the particle, $u$ is its four-velocity, and $\DD /%
\DD\tau $ is covariant differentiation with respect to the proper time $\tau $%
. To obtain this self-force, it is assumed that the power radiated by an
accelerated charge, in its instantaneous rest frame, is
\be \label{23}
P= \frac{e^2}{6\pi} \frac{\dd {\bf v}}{\dd t} \cdot \frac{\dd {\bf v}}{\dd t},
\ee
which is proved by calculating the amount of the electromagnetic energy
escaping to the infinity, for a {\em localised} accelerated charge in the
Minkowski spacetime. Moreover, to obtain (\ref{22}) from (\ref{23}), an
integration by part is needed, which is also valid (the boundary terms
vanish) provided the charge is {\em localised}. Neither (\ref{23}), nor (\ref
{22}) can be proved for non-localised accelerated charges. However, if
one assumes that the self-force experienced by a particle is {\em local},
that is, it depends only on the status of the charge in an infinitesimal
neighbourhood of an instant, one can generalise (\ref{22}) for the case where
the charge is not localised, or the spacetime is not Minkowskian. This does
{\em not} necessarily mean that (\ref{23}), or its relativistic
generalisation
\be
   P={\frac{{e^2}}{{6\pi }}}
   {\frac{{\DD u}}{{\DD\tau }}}\cdot {\frac{{\DD u}}{{\DD\tau }}},
\ee
are also valid in these more general cases.

Now consider the Abraham-Lorentz-Dirac force in two special cases: \\ {\bf A-%
} A charge, the four-velocity of which is proportional to a static time-like
Killing vector field, i.e., a Killing vector field for which there exists a
family of hyper-surfaces normal to it. It is obvious that this situation may
only happen in a static spacetime. In this case, $u^0=1/\sqrt{-\xi\cdot\xi}$%
, and $u^i=0$. Also note that $\dd\left(\xi\cdot\xi\right)/\dd\tau =0$, and
that in a static spacetime $\Gamma^0{}_{00}=\Gamma^i{}_{0j}=0$, which shows
that
\begin{eqnarray}
{{\DD u^0}\over{\DD\tau}}&=&0,\\
{{\DD u^i}\over{\DD\tau}}&=&\Gamma^i_{00}\left( u^0\right)^2,
\end{eqnarray}
and
\be
    {\frac{{\DD^2 u^i}}{{\DD\tau^2}}}=
    \left( u^0\right)^2\Gamma^i{}_{0j} {\frac{{\DD u^j}}{{\DD\tau}}}=0.
\ee
Therefore
\be
   f^\mu_{{\rm ALD}}=0.
\ee
So this charge experiences no self-force, even though its four-acceleration
may be nonzero (if $\xi\cdot\xi$ is space-dependent). In other words, the
force needed to accelerate a charge to this specific four-velocity ($u^0=1/
\sqrt{-\xi\cdot\xi}$, and $u^i=0$) is the same as the force needed to
accelerate an uncharged particle of the same mass. This is in agreement with
the conclusion of section IV: ``An accelerated charge distribution does not
radiate according to certain observers, whenever the charged particles and
the observers move along the same static Killing vector field.'' The case of
a uniformly accelerated charge in the Rindler rocket, discussed in section
III is a special case. \\ {\bf B-} A freely falling charge, in an {\em %
arbitrary} spacetime. In this case, the four-acceleration is zero ($\DD %
u^\mu /\DD\tau=0$) which shows that the self-force is zero. Note that in a
general spacetime, it may be impossible to define an energy-like quantity,
since there may be no time-like Killing vector field. The above conclusion
in a sense shows that a freely falling charge does not radiate; in the sense
that the world-line of a freely falling charged particle is the same as that
of an uncharged particle.
\sect{7. CONCLUSIONS} \nopagebreak
   The notion of radiation in terms of receiving electromagnetic
   energy by an observer is not absolute, but this relative notion
   is consistent with the principle of equivalence.  That is, in a
   static spacetime, a supported charge does not radiate according
   to another supported observer; neither does a freely falling charge
   according to a freely falling observer. Also, a freely falling charge
   does radiate according to a supported observer, and a supported charge
   does radiate according to a freely falling observer.

   The absolute meaning of radiation, i.e. radiation according to world
   line of the charge, was also discussed. We saw that a supported charge
   in a static spacetime, or a freely falling charge, do not radiate, in
   the sense that no extra force is needed to maintain their world-line
   the same as that of a neutral particle.
   \\[\baselineskip]
 {\bf REFERENCES }
\begin{enumerate}
   \item  M. Born, {\em Ann. Physik} {\bf 30}, 1 (1909).
   \item  W. Pauli, in {\em ``Enzyklop\"adie der Mathematischen
                    Wissenschaften''}, Vol. 5, p. 539, esp. p. 647 f.
                    (Teubner, Lipzig, 1918). Translated as W. Pauli,
                    ``Theory of Relativity'', p. 93. (Pergamon,
                    New York, 1958).
   \item  M. v. Laue, ``Relativit\"atstheorie'', 3ed ed, Vol. 1.
                 (Vieweg, Braunschweig, 1919).
   \item  G. A. Schott, ``Electromagnetic Radiation'', p. 63 ff.
   (Cambridge Univ. Press, London, 1912).
   \item  G. A Schott, {\em Phil. Mag.} {\bf 29}, 49 (1915).
   \item  S. M. Milner, {\em Phil. Mag.} {\bf 41}, 405 (1921).
   \item  D. L. Drukey, {\em Phys. Rev.} {\bf 76}, 543 (1949).
   \item  M. Bondi and T. Gold, {\em Proc. Roy. Soc.} {\bf A229},
                 416 (1955).
   \item  T. Fulton and F. R\"ohrlich, {\em Ann. Phys.}
             (NY) {\bf 9}, 499 (1960).
   \item  F. R\"ohrlich, {\em Ann. Phys.} (NY) {\bf 22}, 169
                 (1963).
   \item  A. Kovetz and G. E. Tauber {\em Am. J. Phys.}
                 {\bf 37}, 382 (1969).
   \item  W. Rindler, {\em ``Essential Relativity''}, 2ed edition
   (Springer-Verlag, New York, 1977).
   \item  D. Boulware, {\em Ann. Phys.} (NY) {\bf 124},
                 169 (1980).
   \item  A. Singal, {\em Gen. Rel. Grav.} {\bf 27}, 953 (1995).
   \item  S. Parrott, {\em Gen. Rel. Grav.} {\bf 29}, 1463
                       (1997).
   \item  S. Parrott, {\em ``Radiation from a uniformly
   accelerated charge and the equivalence principle''}, gr-qc/9303025.
   \item  P. A. M. Dirac, {\em Proc. Roy. Soc.} (London)
                    {\bf A 165}, 199 (1938).
   \item  T. W. Darling, F Rossi, G. I. Opat, and G. F. Moorhead, {\em
   Rev. Mod. Phys.} {\bf 64}, 237 (1992).
   \item  J. L. Synge, {\em ``Relativity''}
                    (North-Holland, Amesterdam, 1956).
   \item  A. Gupta and T. Padmanabhan, {\em Phys. Rev. D} {\bf 57 },
        7241--7250  (1998).
\end{enumerate}

\end{document}